\documentclass[aps,prl,twocolumn,showpacs,groupeaddress,useAMS]{revtex4}
\usepackage{graphicx}
\usepackage{natbib}

\usepackage{epsfig,psfig,epstopdf,float,color}
\usepackage{subfigure}
\usepackage{mybibdefs}
\usepackage{multirow}
\usepackage{rotating}

\def\fnl{f_{\rm NL}}

\newcommand{\dd}{\mathrm{d}}

\newcommand{\solM}{\mathrm{M_{\odot}}}
\newcommand{\Mpc}{\mbox{Mpc}}

\newcommand{\h}{h}

\begin{document}

\title{Implications of multiple high-redshift galaxy clusters.}

\author{Ben Hoyle$^{1,2}$, Raul Jimenez$^{1,3}$, Licia Verde$^{1, 3}$}
\affiliation{$^1$  Institute of Sciences of the Cosmos  (ICCUB) and IEEC, Physics Department, University of Barcelona, Barcelona 08024, Spain.\\ $^2$CSIC, Consejo Superior de Investigaciones CientiÞcas, Serrano 117, Madrid, 28006, Spain.\\ $^3$ ICREA, Instituci\'o Catalana de Recerca i Estudis Avan\c{c}at,Passeig Lluis Companys 23, Barcelona 08010, Spain\.\\ }

\date{\today}

\begin{abstract}
To date, $14$ high-redshift ($z>1.0$) galaxy clusters with mass measurements have been observed, spectroscopically confirmed and are reported in the literature. These objects should be exceedingly rare in the standard $\Lambda$CDM model. We conservatively approximate the selection functions of these clusters' parent surveys, and quantify the tension between the abundances of massive clusters as predicted by the standard $\Lambda$CDM  model and the observed ones.  We alleviate the tension considering non-Gaussian  primordial  perturbations of the local type, characterized by the parameter $\fnl$ and  derive constraints  on $\fnl$  arising from the mere existence of these clusters. At the $95\%$ confidence level,  $\fnl>467$  with cosmological parameters fixed to their most likely WMAP5 values, or $\fnl \gtrsim 123$ (at $95\%$ confidence) if we marginalize over WMAP5 parameters priors. In combination with $\fnl$ constraints from Cosmic Microwave Background  and halo bias, this determination implies a scale-dependence of $\fnl$ at $\simeq 3\,\sigma$.   Given the assumptions made  in the analysis, we expect  any future improvements to the  modeling of the non-Gaussian mass function, survey volumes, or selection functions to increase the significance of $\fnl>0$ found here. 
 In order to reconcile  these massive, high-z clusters with an $\fnl=0$, their masses would need to be systematically  lowered by $1.5 \, \sigma$ or the $\sigma_8$ parameter should be $\sim 3 \, \sigma$ higher than CMB  (and large-scale structure) constraints. The existence of these objects is a puzzle: it either  represents a challenge to the $\Lambda$CDM paradigme or it is  an indication that  the mass estimates of clusters is dramatically more uncertain than we think.
 
\end{abstract}
\pacs{cosmology}

\maketitle

\section{Introduction}
Recent developments in observational hardware and observing techniques have enabled the detection of many massive, high-redshift clusters \citep[see, e.g.][]{0607425,0610115,1006.5639,Chiaberge:2010ra}, which seem to create some tension with the abundance predictions of the standard $\Lambda$CDM paradigm \citep{Jimenez:2009us, Holz:2010ck}. Previous work \citep{Matarrese:2000iz,LoVerde:2007ri,D'Amico:2010ta} have examined how the abundance of high-redshift massive clusters within the $\Lambda$CDM model can be enhanced by allowing the primordial fluctuations, a relic of inflation, to deviate from a Gaussian random field.
The most basic models of inflation predict a scale invariant power spectrum of density pertabations $\Phi$, described by a Gaussian random field $\phi$. Probes of the very early Universe \citep[e.g.,][]{Hinshaw:2008kr} and the Large Scale Structure of the late Universe have shown that this description is a good approximation to first order.  However,  any deviations from the slow-roll, single field, adiabatic vacuum state  inflation (and more complex inflationary models) predict deviations from Gaussianity   \citep[see e.g.,][and refs. therein]{Bartolo:2004if, komatsuwhitepaper,Byrnes:2010em}, which are of interest because they 1) Modify the number of high-redshift clusters, relieving tension between theory and observation, and 2) Allow an observational window into early universe physics. The non-Gaussian corrections may  be charaterised by the coefficient $f_{\rm NL}$ \citep[][]{Salopek:1990jq, Ganguietal94, VWHK00, KS01}, which affects the initial potential field $\Phi$, as
\begin{equation}
\Phi=\phi+f_{\rm NL}\left(\phi^2-\langle\phi^2\rangle\right)~.
\label{eq:fnl}
\end{equation}
in the so-called local non-Gaussianity case.

Observations of the Cosmic Microwave Background (CMB) WMAP3  by \citet[][]{Yadav:2007yy}, measured $\fnl$ to be within  $27 < \fnl < 147$ (at the $95\%$ confidence level).  More recently, \citet[][]{Komatsu:2010fb} find $-10 < \fnl < 74$ (at 95\% C.L.), consistent with the above range but also consistent with zero. The CMB constrains $\fnl $ at large scales ($<0.03 \,\h/\Mpc$), but on smaller scales  the Large Scale Structure (LSS) can also constrain $\fnl$ through the clustering \citep[see e.g.,][and refs. therein]{Verde:2010wp,2010MNRAS.407.2339S}  and abundances of massive halos \citep[see e.g.,][]{Matarrese:2000iz,LoVerde:2007ri}. Measurements of $\fnl$ using  LSS, provides complementary constraints to the CMB  and probes any scale dependence of $\fnl$. Considering the   scale-dependence on halo bias induced by local non-Gaussianity, \cite{Xia:2010pe} obtain  $\fnl \sim 53 \pm 25$ at $1\,\sigma$, ($10 < \fnl < 106$ at 95\% confidence) from  the NVSS survey; this signal comes from scales $k \sim 0.03 \,h/\Mpc$.

The detection of the high-redshift cluster of galaxies  {\tt XMMUJ2235.3+2557} \citep{Mullis:2005hp} and a Hubble Space Telescope weak lensing mass measurement \citep{Jee:2009nr}, allowed  \cite{Jimenez:2009us} to show how the tension between $\fnl=0, \, \Lambda$CDM (which predicts $\sim 2 \times 10^{-3}$ such clusters)  and this cluster could be alleviated with values of $150<\fnl<260$.  Massive clusters abundance probes $\fnl$ on scales corresponding to the Lagrangian radius of the halos; $k > 0.1 \,h/\,\Mpc$.

\cite{Holz:2010ck} then calculated at which redshift and mass, the most massive cluster in the Universe was expected to be found, and how this changed with survey volume. They also found that {\tt XMMUJ2235.3+2557} was more than $2\sigma$ away from  $\fnl=0$, $\Lambda$CDM predictions.

Finally, \cite{Cayon:2010mq} formally calculated the constraints which could be placed on $\fnl$ using  {\tt XMMUJ2235.3+2557}. They computed the probability that the ``most massive" cluster expected within the survey volume had a mass, 1) greater than the $68\%$ upper mass estimate of the cluster, 2) within the $68\%$ upper and lower bounds on the mass estimate, and 3) less than the $68\%$ lower bound on the clusters mass. They Poisson sampled from these abundances to obtain a probability that a cluster with the mass of {\tt XMMUJ2235.3+2557} was the ``most massive" system. By exploring how values of $\fnl$ modified cluster abundances \citep[using][]{Matarrese:2000iz}, they placed constraints on $\fnl$ to be greater than zero at the $95\%$ significance level.  We note that $\fnl>0$ is only one possible explanation of the existence of high-redshift massive clusters \citep[see, e.g.,][]{2011MNRAS.tmpL.190B}.

The above studies represent the latest results for constraining $\fnl$ on ($\sim10\Mpc$) cluster scales, and have concentrated on the above single cluster at high-redshift. We extend these previous works by exploring the constraints on $\fnl$ using $14$ high-redshift ($z>1.0$) spectroscopically confirmed galaxy clusters with masses measured in the literature. 

The layout of the paper is thus;  we begin by reviewing the theoretical form of the cluster mass function and the  non-Gaussian correction to it, and continue  by describing the compilation of a high-redshift cluster sample. Here we discuss our conservative assumptions about the selection functions and survey volumes. We then describe our analysis and find the best fitting cosmological parameters, followed by our conclusions and discussions. Throughout the paper, unless otherwise stated,  we assume a flat $\Lambda$CDM model with WMAP5 \citep[][]{Hinshaw:2008kr} cosmological parameters (i.e, $\Omega_m, \, h, \, n_s,\, \sigma_8= 0.28, \, 0.705,\,0.960, \, 0.812$), and quote $\fnl$ using the LSS convention, e.g. $\fnl^{CMB} \simeq \fnl^{LSS}/1.3$ \citep[see, e.g.,][]{Verde:2010wp}.

\section{The non-Gaussian cluster mass function}
\label{theory}
The theoretical cluster mass function was first written down by \cite{1974ApJ...187..425P} who assumed spherically collapsed halos, and was later improved e.g., \cite{Sheth:1999su}. Subsequently, large-volume, high resolution N-body  simulations have been performed and  mass functions fitting formulae have been found \citep[see, e.g.,][]{Jenkins:2000bv,Tinker:2008ff,Bhattacharya:2010wy}.  We use the spherical overdensity Gaussian mass function given by \cite{Jenkins:2000bv}, which determines the number of haloes as a function of mass as measured within a radius at which the density contrast is $180$ times the background matter density  $\rho_m$, and has the form,
\begin{eqnarray}
\label{f_nmz} n(M,z) & = &  \frac{\bar{\rho}}{M} f \Big(-\frac{\dd \ln \sigma_M}{\dd \ln M} \Big) \;, 
\end{eqnarray}  
where $\sigma_M$ is the {\it rms} variation of the density field, smoothed on scales $M$. For a discussion of the minor differences between $180 \rho_m$ and $200 \rho_m$ mass functions see \cite{Tinker:2008ff}. We use the icosmo\footnote{http://www.icosmo.org/} package \citep{Refregier:2008fn} to calculate $\sigma_M(z)$, co-moving distances and other cosmology-dependent parameters, and use the functional form of $f$ \citep[see Equ. B4 of][]{Jenkins:2000bv} given by
\begin{eqnarray}
\label{f_b4} f & = & 0.301 \exp \Big( - | \log\big( \sigma_M(z)^{-1}\big)  +0.64 |^{3.82} \Big) \;.
\end{eqnarray}  

Non-Gaussian corrections to the mass function have been proposed in the literature \citep[][]{Matarrese:2000iz,LoVerde:2007ri, Maggiore:2009rx, D'Amico:2010ta}, and over the mass and redshift ranges considered here, agree to within $10\%$ \citep[see Fig. $5\,\&\,6$ of ][]{D'Amico:2010ta}. These corrections are typically written as the ratio of the non-Gaussian to Gaussian mass functions ${\cal R}$, and are, for example, found by lineararising the $3$-point expansion of the collapse density \citep[as in][]{LoVerde:2007ri}, or by using saddle point approximations to non perturbatively account for higher order corrections \cite[as in][]{Matarrese:2000iz} (MVJ), \citep[although, see][]{Maggiore:2009rx}. We adopt the MVJ prescription to describe how the ratio of the non Gaussian to Gaussian mass functions change as a function of $\fnl$
\begin{eqnarray}
{\cal R}(S_{3,M},M,z) &=& \frac{n(M,z,\fnl)}{n(M,z,\fnl=0)} \; ,
\end{eqnarray}
where $S_{3,M}$ describes the normalized skewness of the smoothed density field, and can be used to define a ``skewness per $\fnl$ unit" as $S_{3,M}= \fnl \;S_{3,M}^{\fnl=1}$. ${\cal R}$ is given by
\begin{eqnarray}
\label{eq:ratioMVJellips}
&&{\cal R}_{NG}(M,z,f_{NL})=  \exp\left[\delta_{ec}^3
\frac{S_{3,M}}{6 \sigma_M^2}\right] \times \\
& &\!\!\!\! \left| \frac{1}{6}
\frac{\delta_{ec}}{\sqrt{1-\frac{\delta_{ec}S_{3,M}}{3}}} 
\frac{dS_{3,M}}{d\ln \sigma_{M}}  
\!+\! \sqrt{1-\frac{\delta_{ec} S_{3,M}}{3}}\right| \; ,
\nonumber 
\end{eqnarray}
where $\delta_{ec}$ is the critical density for ellipsoidal gravitational collapse. \cite{Wagner:2010me}  recently tested these predictions for generic non-Gaussianity, using a suite of N-body simulations, but due to difficulty in computing the initial conditions, they probed relatively low mass ($\le5\times10^{14} \,\solM$) systems. They found that the MVJ mass function may slightly over predict the abundances of massive $\le 5 \times10^{14}\solM$ clusters at high-redshift. If this result can be extrapolated to more massive clusters at even higher redshifts, then the over prediction of the MVJ  non-Gaussian mass function will only strengthen the conclusions drawn from this work, as a larger value of $\fnl$ will be required to fit the observed abundances of massive clusters using a more accurate model, implying this analysis is conservative. 
 
After the publication of this work, \cite{2010arXiv1012.2732E} found that the exponential fall in the \cite{Jenkins:2000bv} mass function is not enough to counter the exponential increase in the non-Gaussian correction  \cite{Matarrese:2000iz}  for very large values of $\fnl$  and large masses ($\gtrsim10^{16}\solM$). They find that the \cite{Tinker:2008ff} mass function is more well behaved for larger values of $\fnl$ and masses, but still breaks down at very large scales. We stopped the mass function integration at $10^{16}\solM$ just before the  \cite{Jenkins:2000bv}  mass function breaks down.  They additionally checked and confirmed the robustness of our method to the choice of the mass function and measure  a mean value  very close to that measured here, for the same sample of clusters, even after correcting for the mass function approximation. In what follows, we only place a lower constraint on the value of $\fnl$, and thus our approach is robust to the choice of mass function at these lower values of $\fnl$ and masses considered.

\section{Data}
\label{data}
\begin{center}
\begin{table*}
  \begin{tabular}{r r r r r} 
Cluster Name &Redshift  &  M$_{200}$ $10^{14}  \solM$ & Method & Mass reference \\ \hline
'WARPSJ1415.1+3612' $^{+}$ & $1.02$ & $3.33^{+2.83}_{-1.80}$ & Velocity dispersion & \cite{0911.0138}\\
'SPT-CLJ2341-5119' $^*$ & $1.03$ & $7.60^{+3.94}_{-3.94}$ & Richness  &  \cite{1003.0005}\\
'XLSSJ022403.9-041328' $^{+}$ & $1.05$ & $1.66^{+1.15}_{-0.38}$ & X-ray & \cite{0709.2300}\\
$\rightarrow$'SPT-CLJ0546-5345' $^*$ & $1.06$ & $10.0^{+6.00}_{-4.00}$ &Velocity dispersion &\cite{1006.5639}\\
'SPT-CLJ2342-5411' $^*$ & $1.08$ & $4.08^{+2.53}_{-2.53}$ &  Richness  & \cite{1003.0005}\\
'RDCSJ0910+5422' $^{+}$ & $1.10$ & $6.28^{+3.70}_{-3.70}$ & X-ray & \cite{0810.1917}\\
'RXJ1053.7+5735(West)' $^{+}$ & $1.14$ & $2.00^{+1.00}_{-0.70}$ &  X-ray & \cite{Stottetal}\\
'XLSSJ022303.0–043622' $^{+}$ & $1.22$ & $1.10^{+0.60}_{-0.40}$ & X-ray  & \cite{Stottetal}\\
'RDCSJ1252.9-2927' $^{+}$ & $1.23$ & $2.00^{+0.50}_{-0.50}$ & X-ray & \cite{0810.1917}\\
'RXJ0849+4452' $^{+}$ & $1.26$ & $3.70^{+1.90}_{-1.90}$ & X-ray &  \cite{0810.1917}\\
'RXJ0848+4453' $^{+}$ & $1.27$ & $1.80^{+1.20}_{-1.20}$ & X-ray & \cite{0810.1917}\\
$\rightarrow$'XMMUJ2235.3+2557' $^{+}$& $1.39$ & $7.70^{+4.40}_{-3.10}$ & X-ray  & \cite{Stottetal}\\
'XMMXCSJ2215.9-1738' $^{+}$ & $1.46$ & $4.10^{+3.40}_{-1.70}$ & X-ray & \cite{Stottetal}\\
'SXDF-XCLJ0218-0510' $^{+}$ & $1.62$ & $0.57^{+0.14}_{-0.14}$ & X-ray & \cite{1004.3606}\\\hline
  \end{tabular}
    \caption{  \label{highzclustable} We compile a list of  high-redshift clusters with mass estimates or measurements from the literature. We show the cluster name, redshift, the mass (converted to $M_{200}$) and $1\,\sigma$ errors, and the mass measurement technique and the mass reference. We mark clusters identified from X-ray surveys by $^+$ and using the SZ SPT survey by $^*$. The $\rightarrow$ indicate the ``least probable" cluster observed in each of the combined surveys.}
\end{table*}
\end{center}
We compile a list of $14$ high-redshift ($z>1.0$) spectroscopically confirmed clusters with masses measured or estimated in the literature, and present them in Table \ref{highzclustable}. We believe this list to represent all known spectroscopically identified clusters with mass measurements. We show the cluster's name, the spectroscopic redshift, the cluster mass and mass error converted to $M_{200}$ (in units of $10^{14} \solM$, assuming an NFW profile \citep[][]{1996ApJ...462..563N} if necessary)  which is the mass enclosed within a radius at which the density is $200$ times that of the background matter density.  and the reference to the mass measurement. We distinguish clusters detected by X-ray surveys and those found using the Sunyaev-ZelÕdovich \citep[][hereafter SZ]{Sunyaev:1972eq} effect. 

Here for each cluster we adopt the mass estimate that gave the least tension (best agreement) with $\fnl=0$ $\Lambda$CDM. For an illustrative example consider two cases; 1) A cluster mass has a large central value ($1\times10^{15} \solM$) with a large error ($4\times10^{14} \solM$) , and 2) a cluster has a slightly lower mass estimate ($7.9\times10^{14} \solM$) with a smaller error bar $9\times10^{13} \solM$) Ê\citep[see][]{1006.5639}.
In our analysis, we find that case 1 is more likely to exist in an $\fnl=0$ $\Lambda$CDM, than case 2. Thus, we use case 1 to be conservative.

We note that mass measurements from different techniques typicaly agree well, e.g. {\tt XMMUJ2235.3+2557} had mass measurements using weak lensing of $8.3^{+2.6}_{-1.9} \times 10^{14}\,\solM$, and $7.3\pm1.3 \times 10^{14}\,\solM$ \citep[][]{Jee:2009nr}, and X-ray mass measurements of $6 \times 10^{14} \, \solM$ \citep[][]{Rosati:2009cm} and  $7.7^{+4.4}_{-3.3}\times10^{14}\,\solM$ \citep{Stottetal}.

We also note that potential high-redshift clusters have been detected, but not followed up spectroscopically \citep[e.g. see][]{Gladders:2004gh}, so their redshifts, and typically, masses are  subject to larger uncertainties, if not unknown. This implies that our analysis can only place a lower limit on $\fnl$ as the other clusters may have higher redshifts and/or be more massive than the clusters in our sample, which would further boost the required value of $\fnl$. 

If any of these potential high-redshift clusters candidates were found to be less massive than those in our sample,  or at lower redshifts,  (and such smaller systems are expected in all $\fnl>0$, $\Lambda$CDM cosmologies), they would not detract from these results using the present selection of clusters, as as  our analysis only consider these ``rare events" that have already been confirmed.

The ability to detect a cluster, measure its redshift and mass for any survey, can be described by the selection function. For believable upper and lower limits to be placed on cosmological parameters (including  $\fnl$) using galaxy clusters, the selection function must be understood. Our analysis uses heterogeneously selected clusters, so combining the selection functions is non trivial. We now describe how we conservatively model the selection functions for the X-ray and SZ surveys. We note that deviations from the conservative modeling, will only strengthen our conclusions.

\subsection{Selection function}
We split the cluster catalogues into two broad categories, those detected using the X-ray by the ROSAT and XMM satellites, and those found using the SZ effect at South Pole Telescope \citep[][hereafter SPT]{Carlstrom:2009um}.

\subsubsection{X-ray}
Many of the X-ray surveys have partially overlapping footprints, differing flux limits and exposure times. This means that  some clusters were multiply detected by distinct groups, e.g. {\tt XMMUJ2235.3+2557} was originally detected by the XMM-Newton Distant Cluster Project  \citep{Mullis:2005hp}, but was later redetected by the XMM Cluster Survey \citep{XCS}. The combination of all of the X-ray surveys,  as performed here, makes the construction of the full survey volume and selection function non-trivial. 

We continue conservatively, by assuming that all X-ray surveys had independent footprints (even if they  did not) and uniform survey volumes (even if some were shallower than others),  which we choose to be between $1.0<z<2.2$ (2.2 represents our estimate of  the deepest survey limit). We find that our conclusions are stable to arbitrary increases of the maximum redshift assumed, but will depend on improvements to the modeling of the survey footprints and volumes. We reiterate that any improvements to the conservative selection function and footprints adopted here, will make any conclusions drawn from this analysis stronger, as a reduced survey volume (caused by a smaller footprint or exposure time), or a worse selection function (i.e. there are clusters in the volume that have not been found) will modify the number of observable clusters expected, which will, at best not change our results, but at worse, increase tension with $\fnl=0$ $\Lambda$CDM. 

The conservative X-ray survey footprint is $294.5$ sq. degrees and is composed of $168$ sq. degrees  from the XMM Cluster Survey, $64$ sq. deg. from the XMM-Large Scale Survey \citep[][]{Pierre:2000ya}, $11$ sq. deg. from the XMM-Newton Distant Cluster Project, $1.3$ sq. deg. from the XMM Contiguous survey \citep[][]{Finoguenov:2009mf}, $17.2$ sq. deg. from the Wide Angle ROSAT Pointed Survey \citep[][]{Perlman:2001th}, and $33$ sq. deg. from the ROSAT Deep survey \citep[][]{Hasinger:1997uv}. 

\subsubsection{SZ}
The SZ SPT survey  has a well understood selection function, and was expected to detect all massive clusters above $ 2 \times10^{14} \, \solM$ \citep[][]{HMH0,Battye:2003bm}, at all redshifts. We again assume a survey volume between $1.0<z<2.2$ and use the footprint of $178$ sq. degrees.  To measure the redshifts of clusters detected with the SZ, one needs optical spectroscopic follow up. Not all the identified clusters have had their redshifts and masses measured \citep[see][]{1003.0005}, but we continue conservatively, by assuming that only clusters with follow-up were detected. This is conservative because future cluster measurements will not relieve the tension with $\fnl=0$ found using the current collection of clusters.
$\,$\\
\begin{figure}
   \centering
\includegraphics[scale=0.4]{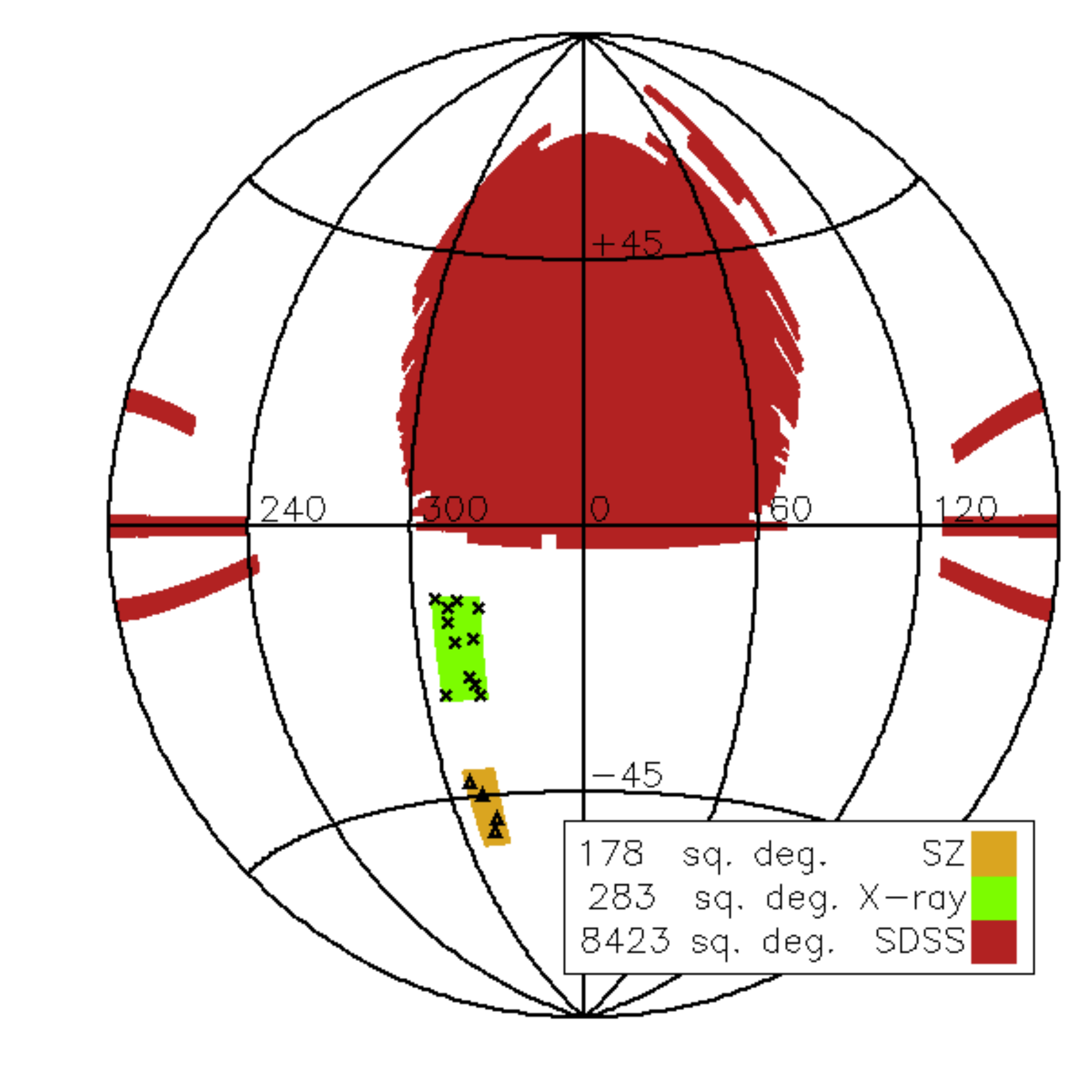}
   \caption{   \label{footprints}An Aitoff projection representing the survey footprints of the combined X-ray survey (shown in green) , the SZ SPT survey (yellow), and we also show the SDSS  survey footprint (red) for comparison.  We note that the X-ray footprints are depicted here as a contiguous region, although the actual footprints consist of pointings across the whole sky. We represent the high-redshift clusters in each survey by crosses and triangles.}
\end{figure}
$\,$
Fig \ref{footprints} is an Aitoff projection representing the survey footprints of the combined X-ray survey (shown as a green contiguous region, although note that the actual X-ray footprint covers many different directions across the full sky), the SZ SPT survey (in yellow), and we also show the Sloan Digital Sky Survey \citep[][hereafter SDSS]{SDSSDR7} survey footprint  for comparison (red). The high-redshift clusters compiled here, are represented by the crosses and triangles. This figure demonstrates how little of the high-redshift sky has been observed, and how much volume remains to find other potentially massive, high-redshift clusters, which may increase the tension with $\fnl=0$ $\Lambda$CDM.

\section{Method and Results}
\label{results}
Our analysis follows two approaches. First we build on the approach of \cite[and refer the reader to \S3 of][]{Cayon:2010mq}, and define the  ``least probable" (i.e., a combination of most massive  and highest redshift) clusters in each of the combined X-ray surveys and the SZ survey, which due to the high mass, should also be the easiest to find.  We then extend the approach of \cite{Cayon:2010mq}, by using the existence of the compiled cluster sample, including  the clusters  full mass error distributions, to examine the probability that the ensemble of clusters could exist in a $\Lambda$CDM universe, and probe how the probability increases with $\fnl$. Initially we keep the cosmological parameters fixed to WMAP5 peak values, and then relax this constraint and marginalize over  WMAP5 priors.

We used the output of both Gaussian ($\fnl=0$) and non-Gaussian (with $\fnl=250$) N-body simulations \citep[obtained from the authors of][]{Wagner:2010me} at a snapshot corresponding to $z=1.0$, to successfully blind test the code pipelines.  We computed the relative values of $\fnl$ needed to explain the existence, abundances, and masses of clusters above $>4\times10^{14}\solM$, after crudely assuming a survey footprint and a redshift slice i.e., a survey geometry. We found that at a fixed ``probability of existing", the recovered value of $\fnl$ for the non-Gaussian simulation data was always $\gtrsim 225$ greater, than that of  the Gaussian simulation data. For the assumed survey geometry we found that the probability of the ensemble of clusters to exist was $40\%$ at $\fnl=0$ in the Gaussian case and $<4\%$ in the non-Gaussian case. In the non-Gaussian case, a value of $\fnl = 230$  is required to obtain a probability of existing  to be $40\%$. We reiterate that the exact recovered probability of existence at fixed $\fnl$ values, depends on the crude conversion of the simulated snapshot volume at $z=1.0$, to the assumed survey geometry, but the differences between the simulations required a value of $\fnl$ similar to that inputted into the non-Gaussian simulations.

\subsection{The least probable clusters}
We begin by asking the question, ``What is the least probable object to be found in each survey assuming $\fnl=0$?". This  approach is  analogous to determining the most massive system in the survey \citep[e.g.][]{Cayon:2010mq}, but generalized to include the redshift-dependence of the mass function. 

Assuming the central value for the clusters mass, we find that the cluster {\tt XMMUJ2235.3+2557} is the least probable X-ray detected object, we expect $5.4$ over the full sky  (and $0.04$ in the X-ray survey footprint)  at $z>1.39$ and $M>7.7\times 10^{14} \,\solM$ using our cosmology and theoretical mass function.  We also find  that {\tt SPT-CLJ0546-5345} is the least probable SZ detected cluster; we expect only $12.5$ over the full sky (and $0.05$ in the survey) with $M>10^{15}\,\solM$ and $z>1.06$.
 
Following \cite{Cayon:2010mq}, we calculate the probability that the mass of the ``least probable" cluster in each survey falls within one of the following three mass bins; 1) less than the $1 \, \sigma$ mass range of the cluster, 2) within the $1\,\sigma$ mass range of the cluster, and 3) greater than the $1 \, \sigma$ mass range of the cluster.  This is accomplished by calculating the theoretical cluster abundance within each mass bin, and then Poisson sampling from these three abundances $10^4$ times (using the same random number seed for each of the three bins), and recording the most massive bin which the Poisson samples is $\ge 1 $. This yields a probability that the ``most massive" cluster exists is within the above mass bins, and within the survey volume.

We then gradually increase $\fnl$, which boosts the abundances of clusters, and Poisson sample from these new abundances to re-derive the above probabilities. This allows us to place constraints on $\fnl$ using the least probable observed cluster in each survey.

\begin{figure}
   \centering
\includegraphics[scale=0.4]{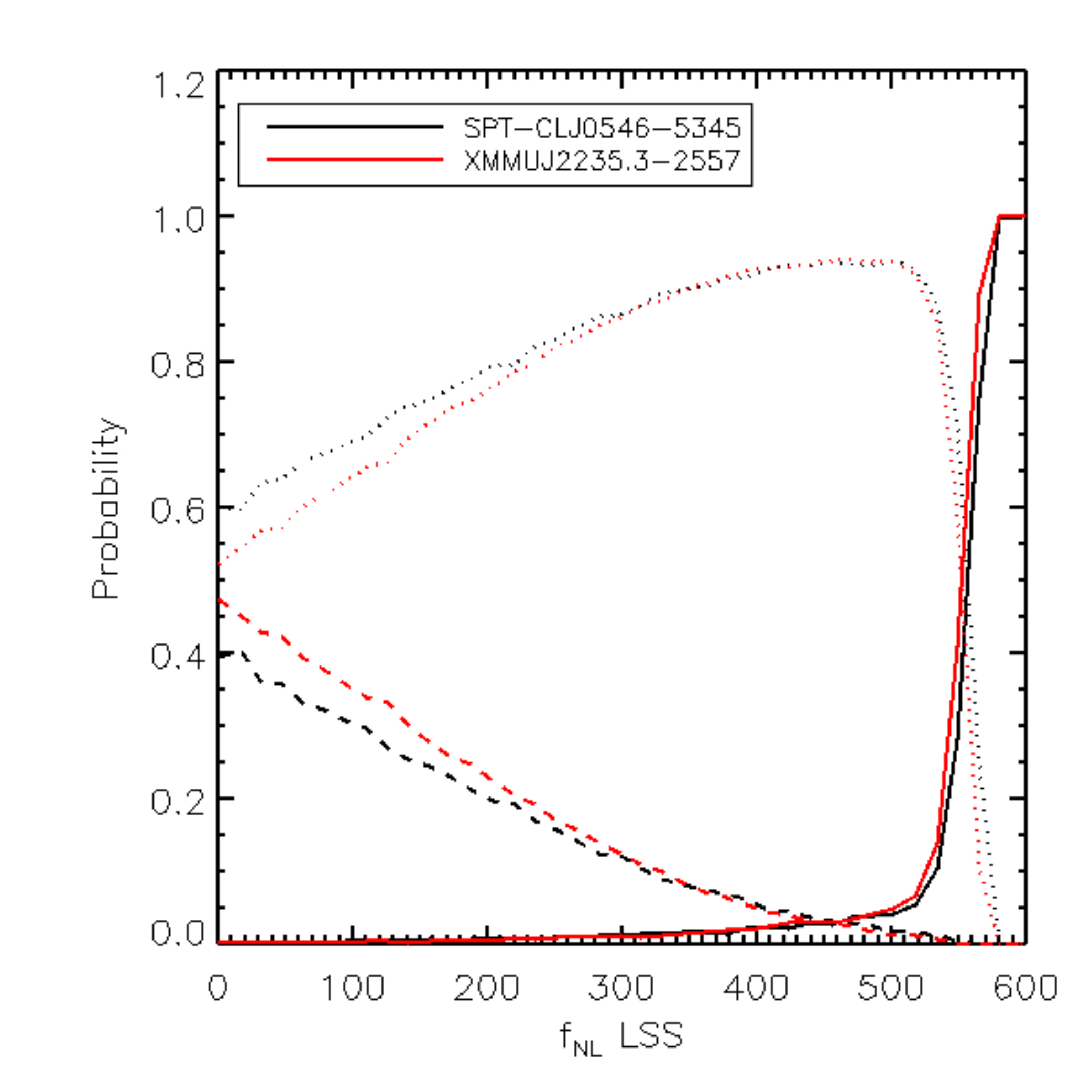}
   \caption{   \label{fnl_leastprob} The probability that the  ``least probable" cluster (in terms of mass and redshift $>z_{clus}$) in each survey could exist, and is the ``most massive" cluster in the surveyed volume, as a function of $\fnl$. The solid line indicates the probability that a cluster more massive than the identified cluster could exist, the dotted line is the probability for a cluster within the measured $68\%$ mass and error range to be the ``most massive system", and the dashed line shows the probability that a cluster less massive than the cluster is the ``most massive" system in the survey volume. }
\end{figure}

In Fig. \ref{fnl_leastprob} we show the probability that each  observed massive cluster is the (theoretically predicted) ``least probable" system in the survey as a function of $\fnl$. We note that both clusters provide similar constraints, which, when combined, points to some tension with $\fnl=0$ $\Lambda$CDM. The constraints obtained here, are slightly different to that in \cite{Cayon:2010mq}, due to differences in the assumed survey footprint, mass function and cosmological parameters. 
Note that for example {\tt XMMUJ2235.3+2557} has another  (weak lensing-based) mass estimate  which has a higher central value and smaller error-bars. This makes our approach conservative.

\subsection{All clusters}
We proceeded by using the existence, masses and full error budgets, of the $14$ clusters in the sample. To model uncertainties we adopt the following Monte-Carlo approach.
We log Gaussian random sample from each cluster's mass and error $10^{4}$ times producing a set of sampled masses $M_S$, and determine how many clusters $N_S(M>M_S,z_{clus}<z<2.2)$ one would expect to find above each sampled mass and above the redshift of the cluster out to edge of survey volume using the mass function expression.
For each of the $10^4$ sampled masses $M_S$, we Poisson sample $P^O$ from the predicted abundances $N_S$, and noted if the Poisson sample $P^O(N_S)\ge1$, i.e. that a cluster more massive than this cluster with a redshift equal to or greater than this cluster could exist.  This formed a probability $P_i$, that each cluster $i$, could exist (marginalized over its mass uncertainty), rather than forming a probability that the cluster is the ``most massive", as above i.e., the probability a cluster exists is $(\#P^O(N_S)\ge1) / 10^{4}$. We then repeated this analysis for each of the clusters and multiplied the probabilities $P_i$ that each cluster could exist in the surveyed region, to produce a combined probability $P(\fnl)=\Pi \,P_i$, that the ensemble of high-redshift clusters could exist in the modeled universe. We increased the value of $\fnl$ and repeated the analysis to produce a probability distribution and stopped the analysis when the $P(\fnl)=1$, i.e, that all the clusters were likely to exist in the cosmological model and survey volumes.

\begin{figure}
   \centering
\includegraphics[scale=0.4]{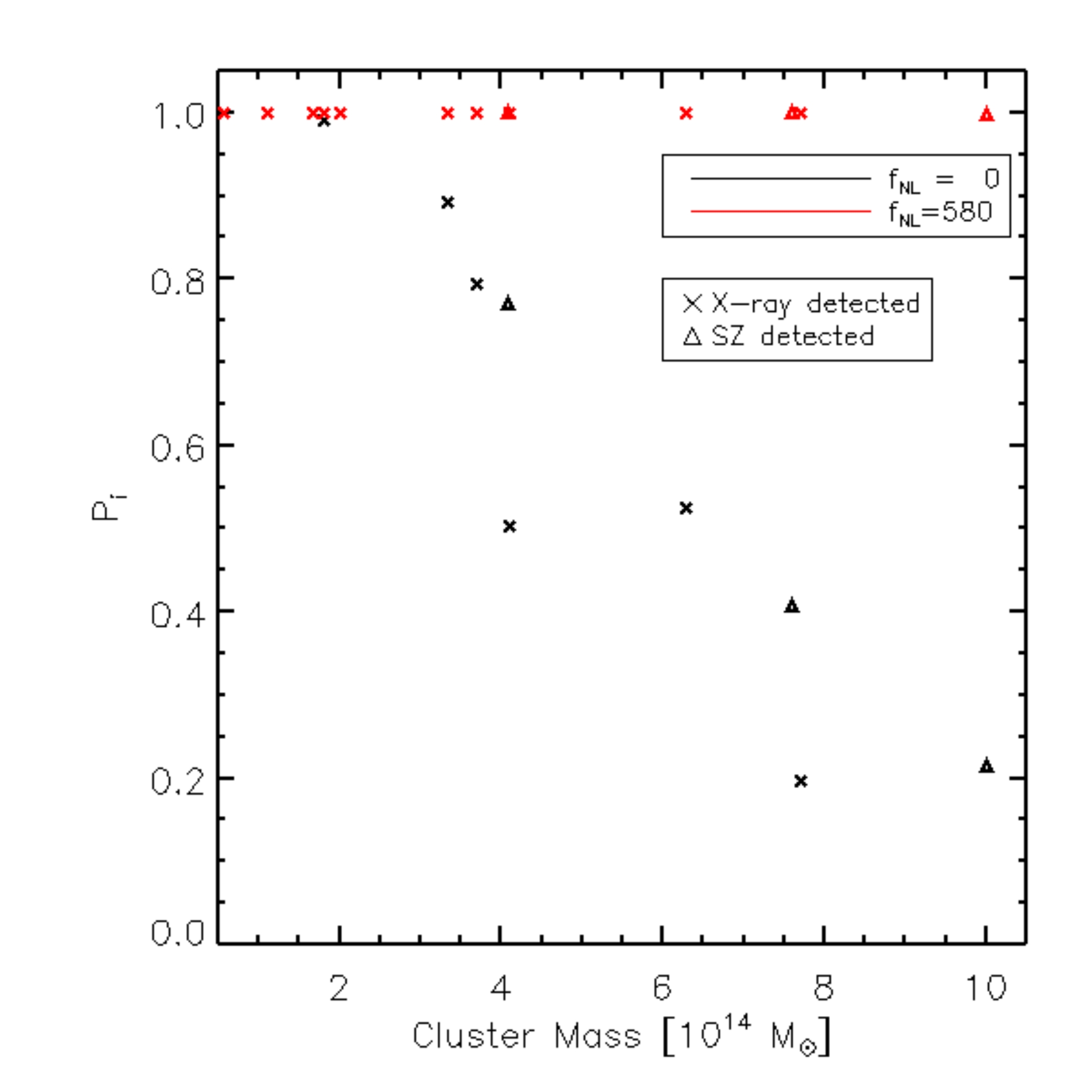}
 \caption{   \label{fnlPall} The probability that each cluster could exist within the survey volume $P_i$, assuming $\Lambda$CDM with $\fnl=0$ (black symbols). We show X-ray identified clusters by crosses and SZ identified clusters by triangles. We also show how the probability changes assuming $\Lambda$CDM and setting $\fnl=580$.}
\end{figure}
Fig. \ref{fnlPall} shows the probability that each cluster could exist given the survey volumes and selection function. The X-ray and SZ identified clusters are distinguished in the figure, but combined in the analysis. We show how the probability $P_i$ for each cluster, varies if we change $\fnl$ from $0$ (black symbols) to $580$ (red symbols).

We see that many clusters are unlikely to exist in a $\fnl=0, \, \Lambda$CDM universe, and by multiplying the probabilities, we find that the probability of the observed Universe being well described by this model is $ 3\times10^{-3}$.  When $\fnl=580$ we note that each cluster is more likely to exist, and the combined probability  $=1$ (for our mass function), which suggests that this model is a better description to the observed Universe \citep[although see,][for a discussion of the validity of the chosen mass function]{2010arXiv1012.2732E}.

\begin{figure}
   \centering
\includegraphics[scale=0.4]{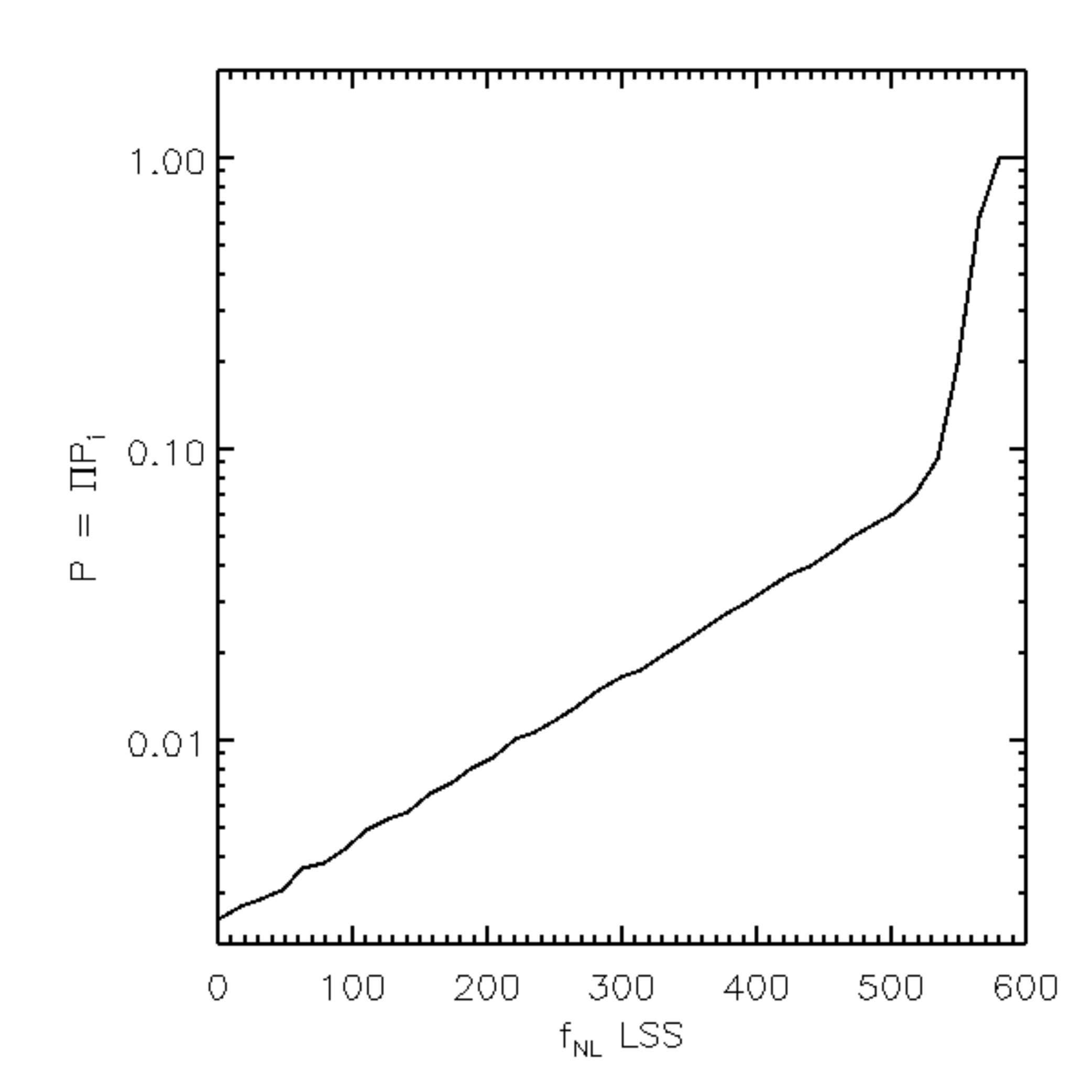}
   \caption{   \label{fnl_all} The probability that the ensemble of clusters could exist in a WMAP5 $\Lambda$CDM universe, as a function of $\fnl$.}
\end{figure}
In Fig. \ref{fnl_all} we plot the combined probability that all the clusters could exist as a function of $\fnl$. We see that the $\fnl=0$ model is a poor fit to the observed Universe, and by increasing $\fnl$ we alleviate tension. We constrain $467<\fnl $ at the $95\%$ confidence level using these clusters. We remind the reader that any improvement in the modeling of the survey volumes, footprints or theoretical mass function, or the detection of more massive, high-redshift clusters, will only increase this result.

\subsubsection{Varying cosmological parameters}
We next simultaneously Gaussian random sample from the parameters $\Omega_M,\, \Omega_{\Lambda},\, \Omega_K\equiv(1-\Omega_M -\Omega_{\Lambda}), \, \Omega_b, \, H_0, \, \sigma_8, \, w_0, \, n_s,$ $\sim1750$ times, using the WMAP5 priors (without imposing spatial flatness) and record the value of $\fnl$ evaluated at $P=0.05$, denoted here as $\fnl |_{P(0.05)}$,  which describes the probability of observing our $14$ clusters $P$ in their surveys $P=0.05$ (i.e. the exsistence of these clusters in their surveys is allowed at $95\%$ C.L.).  This procedure is totally analogous to the so-called ``generalized p-value" for $p=0.05$, where the uncertainty in the clusters mass  and on cosmological parameters is  effectively marginalized over by treating them as ``nuisance parameters" with probability distributions given by the mass estimates and WMAP constraints.

\begin{figure}
   \centering
\includegraphics[scale=0.4]{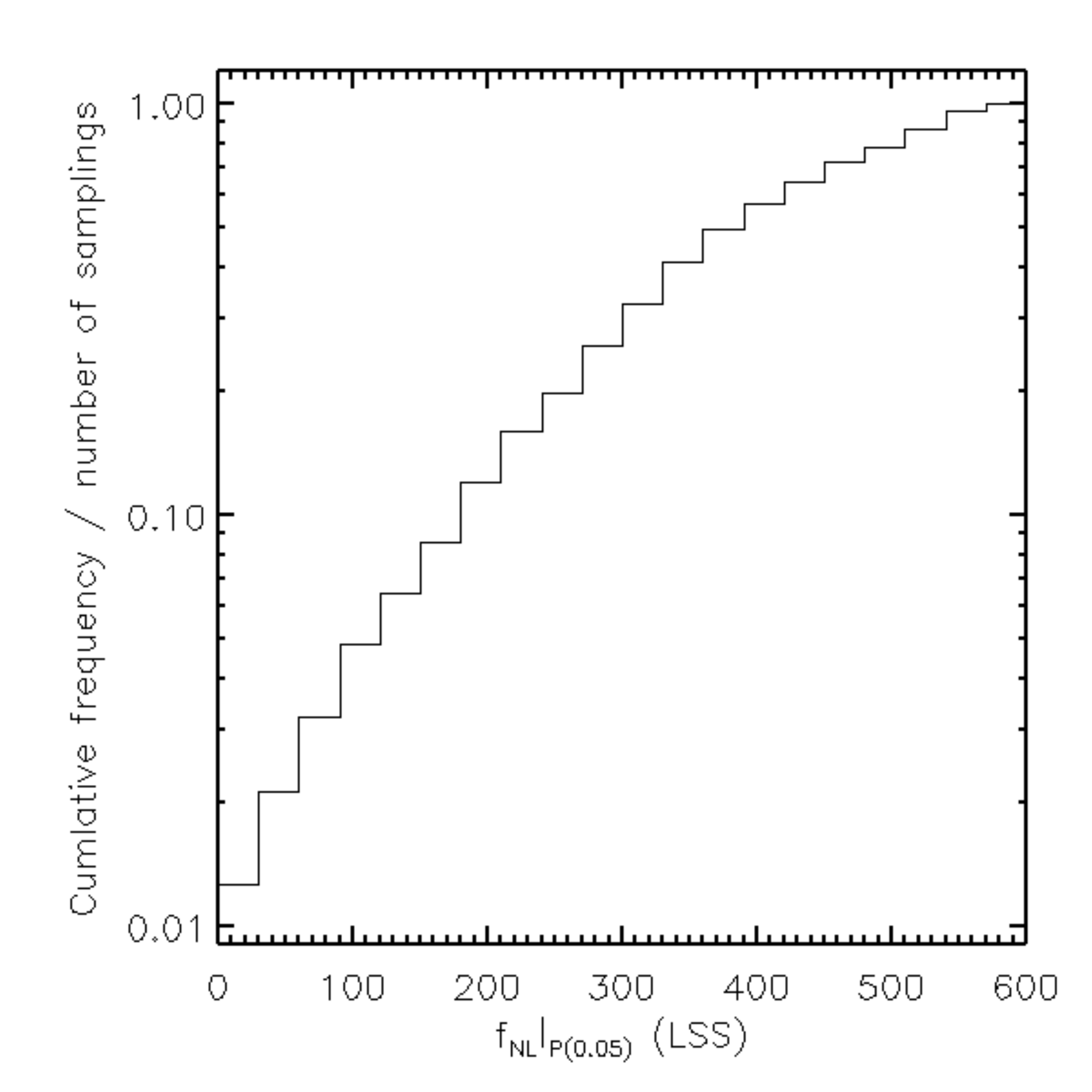}
   \caption{   \label{fnl_cosmo} The distribution of $\fnl$, which correspond to $P(0.05)$ for each Gaussian random sampling of the cosmological parameters $\Omega_M,\, \Omega_{\Lambda},\, \Omega_K, \, \Omega_b, \, H_0, \, \sigma_8, \, w_0, \, n_s$ using WMAP5 priors. }
\end{figure}

In Fig. \ref{fnl_cosmo} we show the $1$d distribution of (generalized) p values (so that $P \ge 0.05$) as a function of $\fnl$. In other words Fig. \ref{fnl_cosmo} shows the frequency in our Monte Carlo procedure of each value of $\fnl$$|_{P(0.05)}$.  We obtain ($123$) $330 <\fnl |_{P(0.05)}$ at $68\%$ ($95\%$) confidence. 

\begin{figure*}
   \centering
   \includegraphics[scale=0.4]{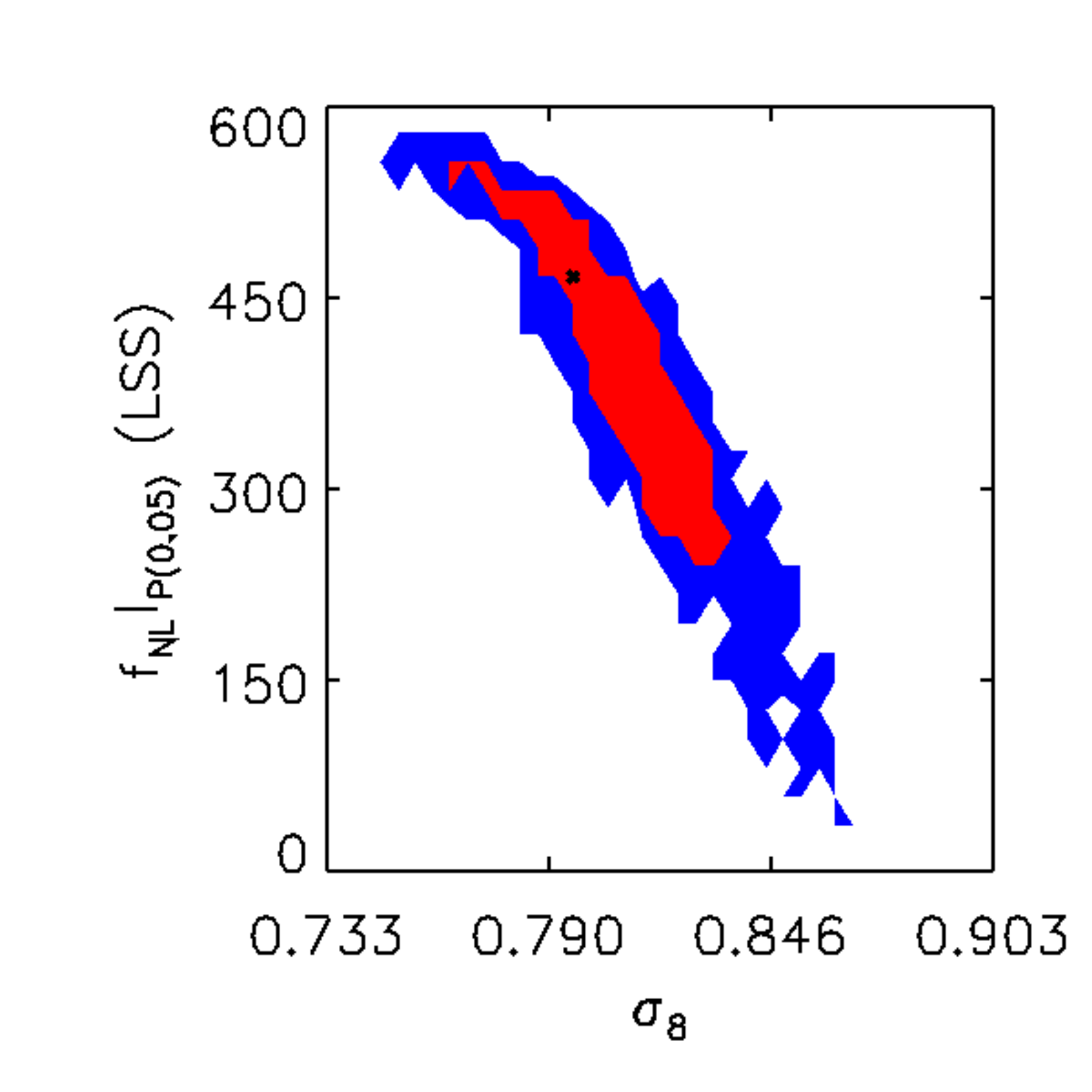}
    \includegraphics[scale=0.4]{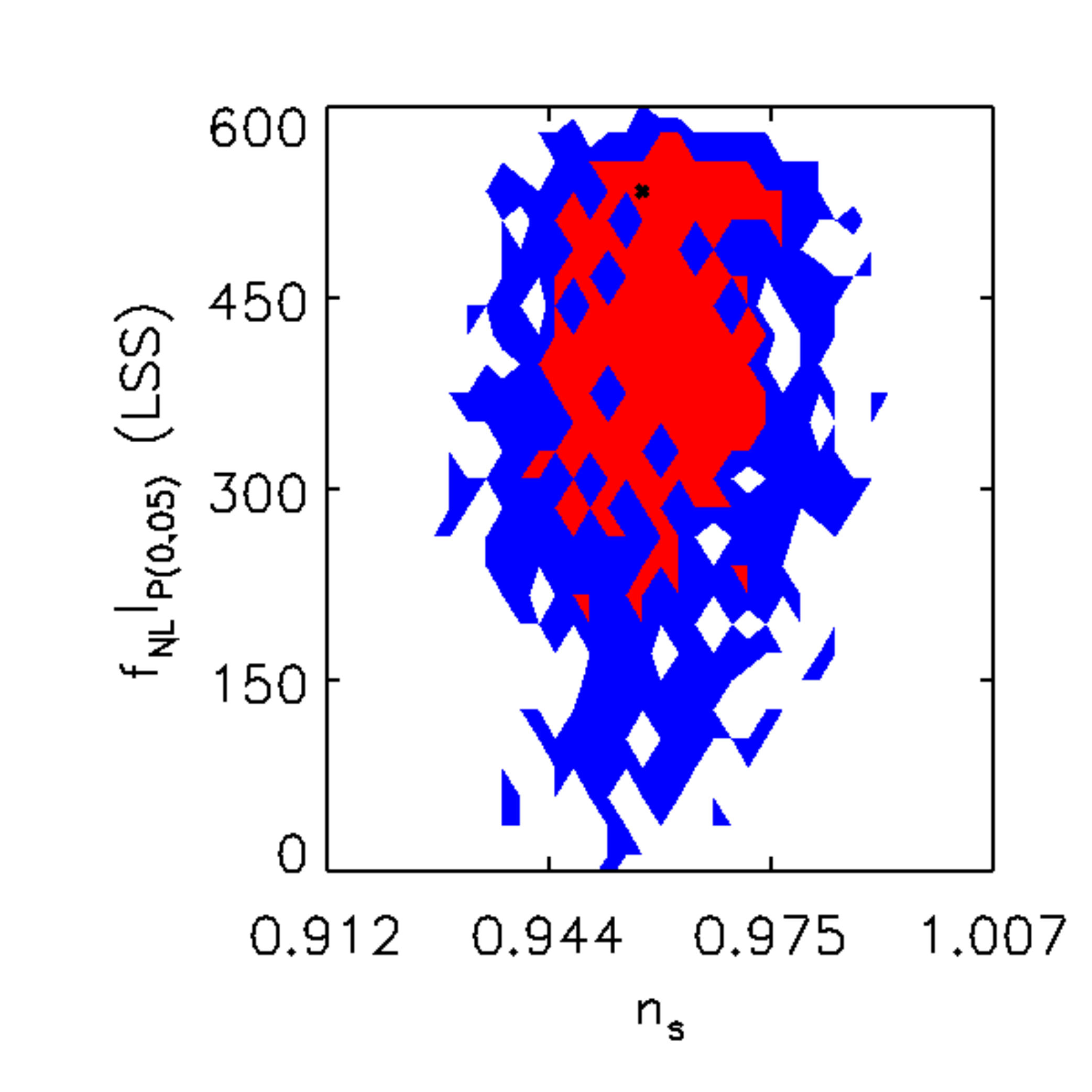}
       \caption{ \label{fnl_2d} Two dimensional marginalized plots for the value of $\fnl$ above which $95\%$ of the probability distribution lies $\fnl |_{P(0.05)}$, against   $\sigma_8$ and  primordial power spectrum spectral index $n_s$. We have represented the peaks in the distributions by thick black crosses and the $66\%$ ($95\%$) confidence levels of these $p$ values by red (blue). Note that these figures represent $p$ value distributions for $p=0.05$, and are not normal MCMC plots. Of all the cosmological parameters explored only $\sigma_8$ shows a degeneracy with $\fnl$.}
\end{figure*}

In Fig. \ref{fnl_2d} we present a selection of two dimensional distributions,  showing the values of $\fnl |_{P(0.05)}$ for the sampled parameter values, against marginalized distributions of;  left) the variance of the density field smoothed on $8\,\Mpc$ scales $\sigma_8$, and  right) the spectral index $n_s$. The filled color contours show the $66\%$ (red) and $95\%$ (blue) significance levels, and we have marked the peaks in each of the distributions by crosses. When viewing these plots, one should keep in mind that they represents $p$ value distributions for $p=0.05$; thus these figures should not be interpreted as standard Markov Chain Monte Carlo plots.

We find that $\fnl$ is degenerate with $\sigma_8$, but less degenerate with all the other varied parameters (we have shown only a selection).  We can calculate the value of $\sigma_8$ needed for  $\fnl$$|_{P(0.05)}=0$ by going to lower $p$ values, or extrapolating down the line of degeneracy using the left panel of Fig. \ref{fnl_2d}, resulting in a value of $\sigma_8 \simeq 0.87$.  If we only vary $\sigma_8$ and keep the other parameters fixed to their WMAP5 peaks values, we find  $\fnl$$|_{P(0.05)}=0$ when $\sigma_8 \simeq 0.89$.

It is interesting to note that  Actacama Cosmology Telescope found   $\sigma_8 < 0.86$ at 95\% CL from upper limits on the SZ power spectrum \citep{FowlerACT}, and SPT found $\sigma_8=0.773\pm 0.025$ \citep{Luekeretal09}. The SZ power spectrum signal depends  very strongly on $\sigma_8$ but not  as strongly on $\fnl$ as, for current observations, it is dominated by massive ($>10^{14} M_{\odot}$) but lower redshift  ($z<1$) clusters (see \cite{Komatsu:2002wc}).  The latest WMAP results  alone (combined with external data sets)  give a more direct, cleaner, measurement  $\sigma_8=0.801\pm 0.03$ ($\sigma_8=0.809\pm 0.024$) \cite{Komatsu:2010fb}.  The high $\sigma_8$ value necessary to obtain $\fnl$$|_{P(0.05)}=0$ is  $\sim 3 \,\sigma$ away from these constraints. 

\section{Conclusions and discussion}
\label{conclusions}
We compiled a list of $14$ high-redshift ($z>1.0$) galaxy clusters with mass measurements from the literature and used their existence to place constraints on the non Gaussianity parameter $\fnl$. The clusters were identified from X-ray surveys and the SZ SPT survey, and we conservatively assumed a selection function and survey volume. We used the theoretical Gaussian mass function of \cite{Jenkins:2000bv} and the prescription for modifying the cluster abundance for non Gaussianities of \cite[][]{Matarrese:2000iz}. We  additionally used the output of the Gaussian and non Gaussian N-body simulations \citep[obtained from the authors of][]{Wagner:2010me} at $z=1.0$, to successfully blind test the code pipelines.

We chose to use cluster mass estimates which were performed assuming a cosmology close to WMAP5 $\Lambda$CDM, and to remain conservative, if more than one measurement technique had been used, we adopted the cluster mass and error measurement which allowed for the lowest sampled cluster mass.

We performed two sets of analysis. First we asked the question, which is the least probable cluster in each survey (this also turns out to be the most massive cluster) and asked how likely this cluster was to be the ``most massive" system in each survey. We found that both massive clusters provide some tension with the $\fnl=0$ WMAP5 $\Lambda$CDM model, and that by multiplying the probabilities, we find that these two clusters have a probability of being observed of $\sim 30\%$.

Using the existence of the $14$ clusters, their masses and full errors distributions, we then calculated the probability that each cluster could exist in the survey. We sampled from each cluster's mass and error and calculating the expected (Jenkins mass function-predicted) abundance above each sampled mass and above the redshift of the cluster, and then Poisson sampled from the abundances\footnote{Subsequently, \cite{2010arXiv1012.2732E} showed that our results are robust to the choice of mass function for the lower bounds placed on $\fnl$ reported here.}. We recorded the frequency that the Poisson sampled number was greater than or equal to one, implying that  at least one cluster with the sampled mass could exist above the redshift of the cluster in the survey volume. We used the frequency of existence to construct a probability that each cluster could exist. We then combined all probabilities, to obtain a final probability that the ensemble of clusters could be found in the modeled universe, and we showed how this probability changes with $\fnl$. We note that our method allows for only a lower limit to be placed on $\fnl$. This is because any new clusters, or improvements to the survey volumes, or selection functions, will increase tension with $\fnl=0$ $\Lambda$CDM with WMAP priors on cosmological parameters. 
 
We found that the best fitting models bound $\fnl$ to be greater than $467$ at the $95\%$ confidence level, when keeping the WMAP5 parameters fixed at their peak values. We also Gaussian random sampled from the cosmological parameters $\Omega_M,\, \Omega_{\Lambda},\, \Omega_K\equiv(1-\Omega_M -\Omega_{\Lambda}), \, \Omega_b, \, H_0, \, \sigma_8, \, w_0, \, n_s$ using the WMAP5 priors. For each realization, we calculated the value of $\fnl$, above which $95\%$ of the probability distribution lay. We find that the median value of $P(0.05)=\fnl$ is $393$, and drops below $\fnl=123$, in only $\sim 5\%$ of realizations. This means that even after marginalizing over cosmological parameters assuming WMAP5 priors, we still find $\fnl |_{P(0.05)}\gtrsim 123$ at the $95\%$ confidence level. 

We have performed several checks: 
{\it i)} the signal is not driven by few objects (e.g., only clusters detected in X-rays or only those detected in SZ, or only clusters which mass estimate is obtained from X-rays etc.)
{\it ii)}  these rare events are not evidently clustered in a special patch of the sky
{\it iii)}  cosmological parameters degeneracies: the $\fnl$ parameter is degenerate only  with the $
\sigma_8$ parameter. To obtain that $\fnl=0$ is allowed at 95\% C. L.,  the value of $\sigma_8$ would have to be $\sim 3\,\sigma$ larger than current cosmological (CMB alone and in combination with LSS) constraints.
{\it iv)} all the cluster mass estimates would have had  to be systematically overestimated by $1.5 \,\sigma$, regardless of the measurement technique used, to allow the ensemble to clusters to be fully compatible with $\fnl=0$ $\Lambda$CDM.

\begin{figure}
   \centering
\includegraphics[scale=0.4]{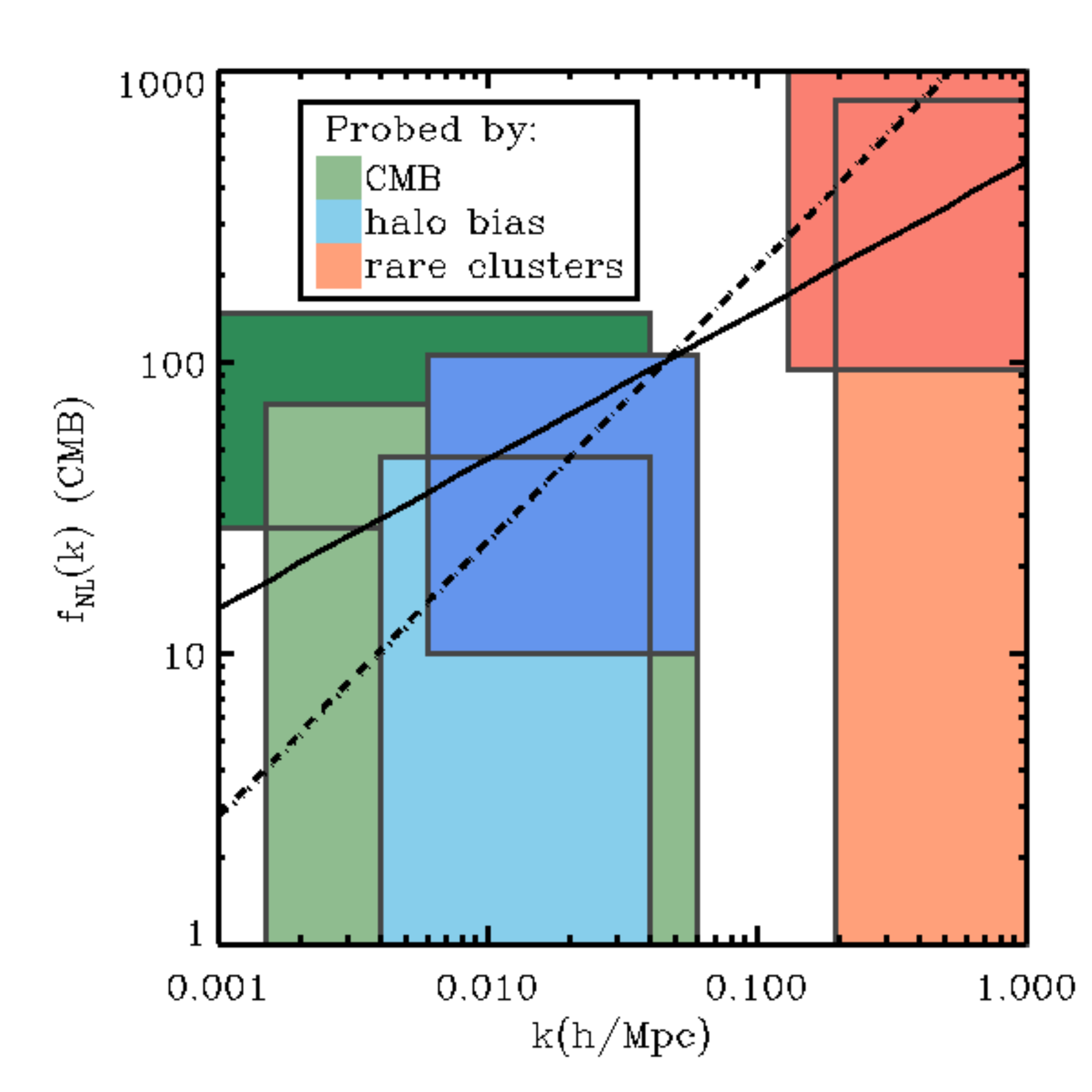}
   \caption{   \label{fnl_licia}A modification of Fig. $8$ of \citet{Verde:2010wp}, with additional  $\fnl$ measurements from the literature (see text) and this work. The colored regions show the scale-dependent measurements of  $\fnl$ using the CMB (green colors), the galaxy halo bias (blue) and the cluster halo abundances (salmon). The enclosed boxes show the $95\%$ confidence levels of each measurement. The small x-axis offsets for different measurements of the same probe is artificial. The lines show the values of scale dependent $\fnl$, see text. }
\end{figure}
In Fig. \ref{fnl_licia}, we compare the result obtained here with other works, using a modified version of Fig. $8$ of \cite{Verde:2010wp}. We overplot the result on CMB scales (using the at $\sim0.04 \, h/\Mpc$ of $27 < \fnl < 147$, at the $95\%$ confidence level by \cite{Yadav:2007yy} (dark green);  of  $\fnl=32 \pm21$ at $1\,\sigma$ by \cite{Komatsu:2010fb} (light green);  the LSS results at scales $\sim 0.4  \, h/\Mpc$ of $449 \pm  286$ at $1\,\sigma$ by \cite{Cayon:2010mq} (light salmon, but note that to apply an upper constraint, they assume that there will be no other clusters found in this footprint as massive or more massive than this cluster); our result of $\fnl^{LSS}>123$  (so $\fnl^{CMB}>95$, dark salmon); the result using a measurement of the non Gaussian scale dependent bias at scales $\sim0.1 \, h/\Mpc$ of $-77<\fnl< 47$ at the $95\%$ C.L. and peaked at $\fnl=8$ by \cite{Slosar:2008hx} (light blue); and the result $\fnl \sim 53 \pm 25$ at $1\,\sigma$  ($10<\fnl<106$ at the 95\% C.L.) by \cite{Xia:2010pe} (dark blue). They also obtained a similar, fully consistent, constraint from the SDSS quasar sample ($\fnl=58\pm 24$). For our application here we use the NVSS numbers.
  
We used these measurements to constrain the non Gaussian spectral index $n_{NG}$, defined by \citep{LoVerde:2007ri}, 
\begin{eqnarray}
\label{fnl_gnl}  \fnl & = &  \fnl^{*} \Big( \frac{k}{k^{*}} \Big)^{n_{NG}} \;,
\end{eqnarray}  
where the $^*$ indicates the CMB pivot scale, $k^*=0.04 \,h/\Mpc$. Note that this scale-dependence parameterization does not allow $\fnl$ to change sign, so in the following approach only $\fnl \ge 0$ is sampled by our procedure. This (theoreticaly-imposed) prior is not too important as $\fnl<0$ for only a small  region with relatively low probability (recall that \cite{Komatsu:2010fb} finds $\fnl>0$ at $1.5$ $\sigma$).

Due to our inability to reliably place an upper constraint on $\fnl$ (see the introduction to the data section for justification), we assumed a log normal distribution for $\fnl^{CMB}$ with a mean of  $5.69$ and $\sigma=0.212$. 

We sampled from the measured values of $\fnl$, while keeping $k$ fixed to the central value, and found the best fitting curve (using MPFIT\footnote{http://cow.physics.wisc.edu/$\sim$craigm/idl/idl.html}) and recorded the value of $n_{NG}$ at each pass. The distribution of $n_{NG}$ is described by $n_{NG}=0.50\pm0.19$  at $1\sigma$, which is a $2.6\,\sigma$ detection of scale dependent bias, using \cite{Yadav:2007yy},  \cite{Slosar:2008hx} and our result,  or $n_{NG}=0.95\pm0.23$ at $1\sigma$,  which is  a $4.0\,\sigma$ detection of scale dependent bias, using \cite{Komatsu:2010fb},  \cite{Xia:2010pe} and our result, or  $n_{NG}=0.93\pm0.23$ at $1\sigma$, using \cite{Komatsu:2010fb}, \cite{Slosar:2008hx} and our result. All of these constraints are in agreement with \cite[][]{Cayon:2010mq}. Since these sets of analysis are not independent, the differing results highlight some possible systematics effects. We show these lines of best fit on Fig. \ref{fnl_licia}.\\

$\,$

For a non flat distribution of objects, each with an observed error, we must account for more objects to be scattered into some part of the distribution than are scattered out. This is described by the Eddington bias, and occurs here because the number of expected very massive clusters above a mass $M$, is exponentially smaller  than the expected number of clusters with mass less than $M$. This could allow lower mass clusters to masquerade as higher mass clusters, and potentially cause us to over estimate $\fnl$.

The Eddington bias is estimated to be only a fraction of the full $1 \sigma$ mass error used in this work, and we have marginalized over the full mass error distribution and have therefore removed any of the Eddington bias effects.  

As a worked example we present the cluster XMMU J2235.3-2557. To calculate the true Eddington bias, one should adopt the more robust cluster mass estimate not, as we have done here, the more conservative one. Typically, the more conservative mass estimate is the one with the largest mass error. E.g., \cite{2010arXiv1011.0004M} states the X-ray mass estimate of  XMMU J2235.3-2557  to be $7.7^{+4.4}_{-3.1}\times 10^{14}\, \solM$.  We find that the statistical correction to the mass $M$, is $\Delta \ln M=0.48$ with $\sigma^2 \ln M=0.16$, and the correction for the Eddington bias is $\Delta \ln M=0.56$, which is indeed higher than the $1 \sigma$ statistical correction (although less than $2 \sigma$). Now, if we instead use the weak lensing mass estimate $M=8.5\pm 1.7 \times10^{14}\, \solM$, of the same cluster, we obtain a statistical correction of $\Delta \ln M=0.2$ with $\sigma^2 \ln M=0.04$, and the corresponding Eddington bias correction is $\Delta \ln M=0.14$. The Eddington bias here is therefore $3.5$ times smaller than the $1 \sigma$ statistical error of the X-ray estimate, which is that used in this work.\\

$\,$
We conclude with the remarks that we have attempted to remain very conservative with our choices of selection functions and volumes, with the cluster mass estimates, and the modeling of the theoretical non Gaussian cluster mass function. Any future improvements in the modeling is expected to strengthen the conclusions of this work; if the survey volume decreases, or more clusters are followed up spectroscopically and found to be massive, or the theoretical non Gaussian mass function modeling is improved, the tension with $\fnl=0$ WMAP5 $\Lambda$CDM will, in all cases, increase.  The existence of high-redshift massive clusters is a puzzle: it  represent a challenge to the $\Lambda$CDM paradigme if the clusters mass estimates reported in the literature  (central values and errors) are taken face value.  These objects grew too massive too fast  compared to the  gravitational instability picture in a $\Lambda$CDM paradigm. Alternatively this is an indication that  mass estimates of high-redshift clusters is dramatically more uncertain than currently believed.  Weak lensing clusters mass estimate is an extremely promising approach to test this possibility as \cite[e.g.,][]{mandelbaum10}
robust and accurate  mass estimates are possible.  Such an observational effort would help address this ``too big, too early"  puzzle.

\section*{Acknowledgments} 
\label{ack}
BH would like to thank Christian Wagner for detailed discussions and making the results of his simulations available, and Shaun Hotchkiss for useful discussions and code comparisons, and LV thanks Carlos Penya Garay for discussions. The authors thank a anonymous referees for comments which improved the paper. BH acknowledges grant number FP7-PEOPLE- 2007- 4-3-IRG n 20218, and the Department of Mathematics and Applied Mathematics at the University of Cape Town for hospitality, LV  and  RJ are supported by MICINN grant AYA2008-0353.  LV is supported by  FP7-IDEAS-Phys.LSS 240117, FP7-PEOPLE-2007-4-3-IRGn202182.
\bibliographystyle{mn2e}
\bibliography{hz}

\end{document}